\documentclass[draft]{svjour3}

\usepackage{natbib}
\usepackage{color}
\usepackage{tikz-cd}
\newcommand\bm[1]{\mbox{\boldmath $#1$}}
\def\npk{\bm{k}}
\def\npl{\bm{l}}
\def\npm{\bm{m}}
\def\npmbar{\bm{\overline{m}}}
\def\half{\smfrac{1}{2}}
\def\quarter{\smfrac{1}{4}}
\def\CKP{Cartan-Karlhede procedure}
\def\d{{\rm d}}
\newcommand{\eqref}[1]{(\ref{#1})}
\newcommand{\smfrac}[2]{{\textstyle{#1\over#2}}}

\makeatletter
\newcommand{\ms}{\noalign{\vspace{3\p@ plus2\p@ minus1\p@}}}
\@addtoreset{equation}{section}
\makeatother

\begin{document}

\title{Spacetimes with continuous linear isotropies III: null
rotations}

\author{M. A. H. MacCallum}
\institute{M. A. H. MacCallum \at School of Mathematical Sciences,
  Queen Mary University of London, Mile End Road, London, UK.
\email{M.A.H.MacCallum@qmul.ac.uk}}

\maketitle

\begin{abstract}
 It is shown that in many cases local null rotation invariance of the
 curvature and its first derivatives is sufficient to ensure there is
 an isometry group $G_r$ with $r\geq 3$ acting on (a neighbourhood of)
 the spacetime and containing a null rotation isotropy. Invariance of
 the second derivatives is additionally required to ensure this
 conclusion in Petrov type N Einstein spacetimes, spacetimes
 containing ``pure radiation'' (a Ricci tensor of Segre type
 [(11,2)]), and conformally flat spacetimes with a Ricci tensor of
 Segre type [1(11,1)] (a ``tachyon fluid'').
\end{abstract}

\section{Introduction}
\label{Intro}

This paper studies local null rotation invariance in spacetime, using
the same definitions, techniques, conventions and notation as the
companion papers studying local spatial rotation invariance and local
boost invariance \citep{Mac21a,Mac21b}, i.e.\ the \CKP\ implemented
using the Newman-Penrose (NP) formalism, the software CLASSI and the
minimal set of totally symmetric spinors defined by
\cite{MacAma86}. Those details will, for brevity, not be repeated
here. As in the previous papers, the ``Newman-Penrose equations'' (the
Ricci equations) and Bianchi identities [(7.21a)-(7.21r) and
  (7.32a)-(7.32k) respectively in \cite{SteKraMac03}] will be referred
to below as (NPa)--(NPr) and (Ba)--(Bk). The original NP notation
$\Lambda=R/24$ will be used, rather than the Ricci scalar $R$, and a
prime on $\Phi_{AB'}$ (denoted $\Phi_{AB}$ in \cite{SteKraMac03}) is
retained for consistency with the compressed notation for symmetrized
spinors representing higher derivatives. Note that the Newman-Penrose
$\Lambda$ and the cosmological constant (typically also denoted by
$\Lambda$) are not the same. Denoting the cosmological constant by $L$
for clarity, $L=R/4$.

Following \cite{Ell67}, (A${}_m$) is the assumption that
\begin{quote}
  At each point
$P$ in an open neighbourhood $U$ of a point $P_o$, there exists a
nondiscrete subgroup $g$ of the Lorentz group in the tangent space
$T_P$ which leaves invariant the curvature tensor and all its
covariant derivatives to the $m$-th order,
\end{quote}
Here, as in the previous papers, the focus is on finding the minimal
$m$ for a given $g$ which ensures that Ellis's definition (C) holds, i.e.
\begin{quote}There exists a local group of motions $G_r$ in an open
neighbourhood $W$ of a point $P_o$ which is multiply transitive on some
$q$[-dimensional] surface through each point $P$ of $W$.
\end{quote}
The $G_r$ will then include, at each $P \in W$, an isotropy subgroup
isomorphic to $g$ leaving $P$ fixed. (In some cases there may in
addition be a proper homothety: see Sections 8.7 and 11.3 of
\cite{SteKraMac03}.)

In this paper the group $g$ is assumed to contain a group of null
rotations\footnote{Here this term denotes one of the transformations
  (3.14) or (3.15) of \cite{SteKraMac03}, rather than the more general
  combination of these with a boost and rotation given the same name
  in \cite{Hal04}. Note that in (3.15) of \cite{SteKraMac03} the final
  $\npl$ should be $\npk$.}.  For such local null rotation invariance,
which will be abbreviated to LNRI, only conformally flat or Petrov
type N spacetimes are possible. For clarity in stating results the
term `null rotation isotropy' will be used only where (C) applies and
the $G_r$ contains, for each $P \in W$, an isometry fixing $P$ which
is a null rotation. This paper aims to find the minimal values of $m$ in (A${}_m$)
for which LNRI implies null rotation isotropy.

Section \ref{prelims} sets out the choices of frame and canonical form
for the curvature in Section \ref{choices}, and their implications,
and then in the following subsections, respectively the conditions
arising from LNRI of a Petrov type N Weyl tensor, from the Bianchi
identities, and from LNRI of $\Phi_{AB'}$.  In Section
\ref{LNRIspaces}, the minimal (A${}_m$) guaranteeing null rotation
isotropy in the different possible types of LNRI spacetimes are
studied. Other than spacetimes with constant curvature, these are
Petrov type N spacetimes with $\Phi_{AB'}=0$ and Petrov type N and
conformally flat spacetimes with non-zero $\Phi_{AB'}$ of certain
Segre types. The appendices summarize the known spacetimes admitting
LNRI and the form of the Bianchi identities with the specializations
of Section \ref{choices}.

\section{Preliminaries}
\label{prelims}

\subsection{Canonical forms of the curvature tensor, and frame choices}
\label{choices}

\subsubsection{The Weyl tensor}
In Petrov type N, the Weyl tensor is invariant under a two-dimensional
group of null rotations.  A suitable canonical form, if the repeated principal
null direction preserved by the null rotations is $\npk$, is one in
which the Weyl spinor has only $\Psi_4 \neq 0$.

In terms of an aligned spinor basis, the group of null rotations fixing
$\npk$ is
\begin{equation}\label{nulrot}
  o^{\star}= o, \qquad \iota^{\star}=\iota+Bo,
\end{equation}
where $\star$ denotes the transformed values and $B$ is a complex
number. If there is invariance of the curvature and its derivatives
for all values of $B$ in \eqref{nulrot}, the LNRI group is
two-dimensional. The transforms of the curvature spinors and their
derivatives will be polynomials in $B$ and $\bar{B}$ with the
untransformed values as coefficients, rather than just multiples of
the untransformed values as in spatial rotation or boost invariance.
The test for LNRI then involves a set of linear equations in
components of the curvature and its derivatives, as in \cite{Mac20},
rather than just the vanishing of some components.

Applying a position-dependent boost\footnote{As in the previous
  papers, choosing specific frames is for computational convenience. The
  results, being invariantly expressed, would be the same in any
  frame, though they might be harder to check.}, $|\Psi_4|$ will be
set to 1 in Petrov type N. A position-dependent spatial rotation will
then be used to make $\Psi_4$ constant: the remaining frame freedom
now consists of a constant spatial rotation and position-dependent
null rotation(s) about $\npk$. The spatial rotation freedom could be
used to set $\Psi_4=1$, but when there is only a one-parameter LNRI
group it is more convenient to use it to
make the parameter $B$ in \eqref{nulrot} pure imaginary.

That is possible, assuming that the one-parameter LNRI group acts in
the same way at all points in the neighbourhood $U$, because in that
case $B=|B|e^{i\theta}$ with a fixed $\theta$ and varying $|B|$. The
remaining rotation freedom can be used to set
$\theta=\pi/2$. (The rotation that does this will not in general also
give $\Psi_4=1$.) With this one-parameter LNRI group there is still the
freedom of a null rotation with position-dependent $|B|$.

In conformally flat spaces $\Psi_4=0$ and the Weyl tensor gives no
information on frame choice. $\Lambda$ also cannot restrict the frame.
The tracefree part of the Ricci tensor, $\Phi_{AB'}$, remains to be
considered.

\subsubsection{The Ricci tensor}

As shown  in \cite{Mac20}, a totally symmetrized spinor
$\chi_{ab'}$ of valence $(m,\,n)$ with $m \geq n$, such as one of the
minimal set of Cartan invariants defined by \cite{MacAma86}, is
invariant under \eqref{nulrot} only if\footnote{The example found by \cite{MilPel09} which requires
  the 7th derivative of the Riemann tensor to complete the \CKP\ is a
  Petrov type N spacetime which has no isotropies.
% (the $s$ sequence starts (2,1,0)).
  However, it obeys the conditions
  \eqref{zerocond}--\eqref{linearcond} at steps 0 and 1 of the
  \CKP\ ({\AA}man, private communication).}
\begin{eqnarray}
\label{zerocond}
\chi_{ab'}&=& 0 \qquad  {\rm for\ all\ pairs}~a+b\leq m-1,\\
\label{linearcond} 0&=& (a+1) B \chi_{ab'}
+b \bar{B} \chi_{(a+1)(b-1)'}~ {\rm otherwise}.
\end{eqnarray}
(Here one can exchange the roles of the pair $m,a$ with those of
$n,b$.) For a
two-parameter LNRI group the only non-zero component is
$\chi_{mn'}$. Note that the conditions for a one-parameter LNRI group
still apply (trivially) in that case.

Assuming a one-parameter LNRI group with $B$ imaginary,
\eqref{linearcond} becomes
\begin{equation}\label{linearcond1}
  (a+1)\chi_{ab'} = b\chi_{(a+1)(b-1)'}.
\end{equation}

For LNRI of the Ricci curvature, $\Phi_{AB'}$ must be invariant under
\eqref{nulrot}.  The Segre types of the possible such Ricci tensors
are listed in Table 1, taken from Table 5.2 of \cite{SteKraMac03}.
\begin{table}[ht] \tabcolsep0.7ex
  \centering\caption{Nontrivial invariance groups containing null
    rotations, by Ricci tensor type.}
  \vspace{2ex}
\label{tab:5.2}
\begin{tabular}{@{}ll@{}} \ms \hline\hline \ms
Invariance group & Segre type of the Ricci tensor \\ \ms\hline \ms
\ms
One-parameter group of null rotations & $[1(1,2)],[(1,3)]$ \\
Null rotations and spatial rotations & $[(11,2)]$ \\
$SO(2,1)$: three-dimensional Lorentz group & $[1(11,1)]$ \\
Full Lorentz group & $[(111,1)]$ \\ \ms
\hline\hline
\end{tabular}
\end{table}

Segre type $[(111,1)]$ implies that $\Phi_{AB'}=0$ for all $A$ and
$B$, so the spacetime is an Einstein space. The Bianchi identities (see
Appendix \ref{bianchi}) then imply
that $\Lambda$ is constant. If such a spacetime is
conformally flat, it is one of the well-known constant curvature
spacetimes, which need not be considered further ((A${}_0$) with $s_0=6$
is sufficient in this case to ensure (C) above). Petrov type N
Einstein spaces are included in Section \ref{Nvac}.

In general relativity, type $[(11,2)]$ can represent ``pure
radiation'', including a null electromagnetic field, and type
$[1(11,1)]$ a tachyonic fluid, but types $[1(1,2)]$ and $[(1,3)]$ have
no generally-accepted physical interpretation, although type
$[1(1,2)]$ could be considered a superposition of pure radiation and a
tachyonic fluid.  Types $[1(1,2)]$ and $[(1,3)]$ were therefore not
considered in \cite{SteKraMac03}, but are retained here for
completeness although, without a physical basis for the form or
values of the Ricci tensor, there are no equations from which
physically meaningful general-relativistic solutions could be
obtained. With any $\Phi_{AB'}$, there could be a nonzero $\Lambda$.

Applying \eqref{zerocond}--\eqref{linearcond}, one finds from
\eqref{zerocond} that $\Phi_{00'}=\Phi_{01'}=0$ and, from
\eqref{linearcond} with imaginary $B$, that $\Phi_{02'} =2 \Phi_{11'}$
if $\Phi_{11'}\neq 0$, and $\Phi_{12'}$ is real if nonzero. When
$\Phi_{12'}\neq 0 = \Phi_{11}$, a position-dependent null rotation can
be used to bring the curvature to a canonical form in which
$\Phi_{22'}=0$. Otherwise $\Phi_{22'}\neq 0$ is possible. These
considerations lead to the following canonical forms in an NP tetrad
with $\npk$ as the null vector preserved by an at least
one-dimensional LNRI group and a compatible Segre type. The nonzero
components $\Phi_{AB'}$ are:\\
{\bf Segre type [(1,3)]}: $\Phi_{12'}\neq 0$ is real.\\
{\bf Segre type [1(1,2)]}: $\Phi_{02'}=2\Phi_{11'}\neq 0 \neq
\Phi_{22'}$.\\
{\bf Segre type [(11,2)]}: $\Phi_{22'}\neq 0$.\\
{\bf Segre type [1(11,1)]}: $\Phi_{02'}=2\Phi_{11'}\neq 0$.\\
The last two cases are specializations of the canonical form for Segre
type $[1(1,2)]$ and there are thus strong correspondences between the
calculations for these three cases.

\subsubsection{Possible LNRI groups and further specializations}
\label{sec:possLNRI}
In conformally flat spacetimes, the null direction $\npk$ defining the
LNRI group cannot be fixed by the (zero) Weyl tensor, and must be
determined using the Ricci tensor and its derivatives. As discussed
above Ricci tensors of Segre type $[(111,1)]$ give only spaces of
constant curvature and Petrov type N Einstein spaces.  A Ricci tensor
of Segre type [1(11,1)] is invariant under null rotations about either
of two directions, say $\npk$ and $\npl$, but if both these are in $g$,
so is a boost invariance in the $(\npk,\,\npl)$ plane because there is
no two-parameter subgroup of the Lorentz group generated by such a
pair of null rotations (see Table 6.1 of \cite{Hal04}). Such boost
invariant spacetimes are conformally flat, have $s=3$, admit a $G_6$
or $G_7$ and were considered in \cite{Mac21b}. Hence for Segre type
[1(11,1)] only spacetimes which are LNRI about $\npk$ and not $\npl$ need be
considered here. The remaining possible Segre types define the $\npk$
direction uniquely.

The canonical forms above have been chosen so that for those
spacetimes admitting only a one-dimensional LNRI group, $B$ in
\eqref{nulrot} is imaginary. Having fixed the $\npk$ direction one may
still apply a spatial rotation and/or boost preserving it. A Ricci tensor of
Segre type [(11,2)] has spatial rotation invariance, but the boost can
be used in the conformally flat case to set $|\Phi_{22'}|=1$, and this
can similarly be achieved in type [1(1,2)]. In conformally
flat spacetimes with a Ricci tensor of Segre type [(1,3)] one can set
$\Phi_{12'}=1$, using the boost and spatial rotation freedoms (a
position-dependent rotation was used to make $B$ imaginary but
the remaining constant rotation freedom can be used to make the necessarily real
$\Phi_{12'}$ positive).

One may note that with these choices of canonical frame in a
conformally flat spacetime the left side of each of the Bianchi
identities is real, so that the imaginary part of each right side must
vanish. Moreover, if $t_0=0$ all components of the curvature are
constants and therefore all the left sides of the Bianchi identities
are zero. The deductions from Bianchi and Ricci identities for
conformally flat spacetimes with Ricci tensors of Segre types [(11,2)]
and [1(11,1)] considered in, respectively, \cite{Mac21a} and
\cite{Mac21b} remain true in the LNRI cases.

\subsubsection{Implications for the \CKP}
\label{sec:LNRIImplns}
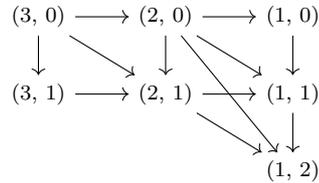
\begin{figure}[ht]
  \caption{Possible pairs $(s,\,t)$ and their evolution in successive
    steps of the \CKP.}
\label{eq:diagram}
% \raisebox{-.5ex}{$1 \rightarrow \mathbb{Z}_2 \rightarrow$}
\begin{center}
\begin{tikzcd}
  (3,\,0)\arrow[r]\arrow[rd]\arrow[d] &
       (2,\,0) \arrow[r]\arrow[d]\arrow[rd]\arrow[ddr] & (1,\,0)\arrow[d] \\
  (3,\,1)\arrow[r] & (2,\,1) \arrow[r]\arrow[rd] & (1,\,1) \arrow[d]\\
            & & (1,\,2)
\end{tikzcd}
\end{center}
\end{figure}

The figure in \cite{Mac21a} which shows possible evolution of pairs
$(s_i,\,t_i)$ at step $i$ of the \CKP, is repeated here as Figure
\ref{eq:diagram}: for full explanation see Section 2.2 of \cite{Mac21a}.
\newline Ricci tensors of types [(11,2)] and [1(11,1)] are invariant
under a three-parameter linear isotropy group, so conformally flat
spaces with such Ricci tensors have as their $(s_0,\,t_0)$ pair one of
the $(3,\,T)$ pairs in Figure 1. However, as discussed above,
spacetimes with a Ricci tensor of type [1(11,1)] either have a boost
invariance and were covered in \cite{Mac21b} or have only a
one-dimensional LNRI group. Similarly spacetimes with a Ricci tensor
of type [(11,2)] which have a spatial rotation invariance were
discussed in \cite{Mac21a}. So in this paper only groups $g$
consisting solely of null rotations need be considered.

All Petrov type N spacetimes with LNRI have $s_0=2$ or 1 and a $g$
consisting only of null rotations.  For LNRI conformally flat
spacetimes the group $g$ of null rotations can be two-dimensional
if $\Phi_{AB'}=0$, which would only give only the constant curvature
spacetimes, or if $\Phi_{AB'}$ is of Segre type [(11,2)]. Otherwise
$s=s_1=1$, so that $(s_1,\,t_1)$ is of the form $(1,\,T)$
where $T$ represents an allowed value of $t$ in Figure
\ref{eq:diagram}.

If $Q$ is a scalar function of position, then the vector
 defined by $Q_{,a}$ is
\begin{equation}\label{NPderivs}
  Q_{,a}(2m^{(a\!\phantom{)}}\bar{m}^{\phantom{(}\!b)}
 -2k^{(a\!\phantom{)}}l^{\phantom{(}\!b)}) = \delta Q\,
 \npmbar+\bar{\delta}Q\, \npm - DQ\, \npl - \Delta Q\, \npk,
\end{equation}
and this must be a linear combination of vectors invariant under the
LNRI group. The number of such independent vectors bounds $t_i$ above.

If there is a a two-dimensional LNRI group, $\npk$ defines the only
direction invariant under $g$, but if the LNRI group is
one-dimensional there are two invariant directions, the second being
given by $B\npmbar-\bar{B}\npm$, which for imaginary $B$ is in the
direction $\npm+\npmbar$.  Whether $s=1$ or 2, \eqref{NPderivs} shows
that for any LNRI scalar $Q$, $DQ=0$, and if $s=2$, $\delta Q =
\bar{\delta}Q=0$ also. If $s=1$ and $B$ is imaginary, $\delta
Q=\bar{\delta} Q$, and this is real for real $Q$. So in either case
one will have, for an invariant $Q$,
\begin{equation}\label{InvtDerivs}
  DQ=0, \qquad  \delta Q=\bar{\delta} Q,
\end{equation}
where $\delta Q$ will be real if $Q$ is, and will be zero if $s=2$.

In the \CKP, this implies that if $s_{i+1}=2$, $t_i \leq 1$. So when
$s=2$ the \CKP\ terminates at the latest at $p=2$ (if the $(s,\,t)$
sequence is $(3,\,0)\rightarrow (2,\,0) \rightarrow (2,\,1)\rightarrow
(2,\,1)$) and thus (A${}_3$) suffices.  If for all $i \geq 0$,
$s_{i+1}=1=s$, then $t_i \leq 2$.  In principle (A${}_4$) could now be
required, for a sequence starting at $(s_0,\,0)$ for some $s_0>1$ and
then running through each of the $(1,\,T)$ possibilities in Figure
\ref{eq:diagram} (so $t_0=t_1=0$). However, this could only happen in
Petrov type N Einstein spacetimes, type N ``pure radiation''
spacetimes, or conformally flat spacetimes with Ricci tensors of Segre
types [(11,2)] or [1(11,1)]. Otherwise, (A${}_3$) would again
suffice. In the sequel it is found that in all those cases
(A${}_4$) is satisfied if (A${}_0$)--(A${}_3$) are, and no new
information arises at the third derivative.

With a $g$ consisting solely of null rotations, if $s=2$ then $t_p\leq
1$ and there will be an isometry group $G_r$ with $r \geq 5$.  If $s=
1$ there is an isometry group $G_3$ with $r \geq 3$. Appendix
\ref{metrics} reviews known spacetime metrics consistent with
these bounds.

After studying the possible combinations of Weyl and Ricci tensors, it
turned out to be more convenient to consider the conformally flat and
Petrov N cases together in most Segre types with $\Phi_{AB'}\neq 0$,
using the Bianchi identities and the conditions imposed on $\Phi$ by
LNRI before those arising from LNRI of $\Psi$. The exception is Segre
type [(11,2)] where the calculations for the Petrov N and conformally
flat cases have fewer common features.

\subsubsection{Additional invariants}
\label{sec:AddedInvts}
The directional derivative conditions above for LNRI apply not only to
components of the curvature and its derivatives in the
canonically-chosen frames, but also to those spin coefficients, or
combinations thereof, completely determined by the frame
specializations. Several cases below lead to
\begin{equation}\label{SCinvce}
  \kappa=\rho=\sigma=\varepsilon=0, \qquad \alpha=\beta+\tau,
\end{equation}
and if that is true the following are invariant under the remaining
frame freedom: $\alpha$, $\beta$, $\tau$, and, when $B$ is imaginary,
$\pi+\bar{\pi}$. When \eqref{SCinvce} holds, if $\alpha=0$ or if there
is a one-parameter LNRI group with imaginary $B$, then $\gamma$ is
also null rotation invariant. For brevity such null rotation
invariants will often be called just invariants in this paper.

Two more invariant combinations of spin coefficients are found in the
cases studied below.  For a one-parameter LNRI group with imaginary
$B$, $\lambda+\bar{\mu}$ is invariant if $\kappa+\bar{\kappa}=0=
\rho+\bar{\sigma}$ and
\begin{equation}\label{lmcond}
   \pi+\bar{\pi}+2(\alpha-\bar{\beta})=0.
\end{equation}
Under the same conditions, $\nu+\bar{\nu}$ is invariant if
$\varepsilon$ is real,
\begin{equation}\label{ptident}
  (\tau+\bar{\tau}) + (\pi+\bar{\pi})=0,
\end{equation}
and
\begin{equation}\label{glmident}
  \lambda+\bar{\mu}+2\bar{\gamma}=2\gamma+ \bar{\lambda}+\mu.
\end{equation}
(i.e.\  $\lambda+\bar{\mu}+2\bar{\gamma}$ is real).
(The invariants' properties $DQ=0=(\delta-\bar{\delta})Q$ can usually
be obtained from the NP equations also, but this has not been checked
in all cases.)

\subsection{LNRI and derivatives of $\Psi$}
\label{AmWconds}

% File typen.nul
In a canonical frame chosen as above, all directional derivatives of
$\Psi_4$ vanish and $\nabla\Psi_{AB'}$ has as possibly non-zero
components
\begin{eqnarray}
\nabla\Psi_{30'} &=& 2\kappa\Psi_4/5, \qquad \nabla\Psi_{31'} =
2\sigma\Psi_4/5,\nonumber \\
\nabla\Psi_{40'} &=&
   [4(\varepsilon+\rho)\Psi_4]/5, \nonumber \\
\nabla\Psi_{41'} &=&
   [4(\beta+\tau)\Psi_4]/5, \label{DPsi}\\
   \label{DPsi2}
\nabla\Psi_{50'} &=& 4\alpha\Psi_4, \qquad
   \nabla\Psi_{51'} =4\gamma\Psi_4\nonumber.
\end{eqnarray}

Whether the LNRI group has $s=1$ or 2, if $\Psi_4\neq 0$,
\eqref{zerocond} and
\eqref{linearcond} imply
$\nabla\Psi_{30'}=\nabla\Psi_{31'}=\nabla\Psi_{40'}=0$, so that
\begin{equation}
\label{Nconds1}
\kappa=\sigma=\varepsilon+\rho=0.
\end{equation}
The Ricci tensor has not yet been specified, but one may note that the
null rotation invariance ensures that condition (C) of the
Kundt-Thompson theorem as stated in \cite{SteKraMac03} is satisfied,
and also that when $\rho=0$ these spacetimes are members of Kundt's
class (Chapter 31 in \cite{SteKraMac03}). \eqref{Nconds1} implies that
in both the Petrov N and conformally flat cases terms $\sigma\Psi_4$
and $\kappa\Psi_4$ which would otherwise appear in (Bg) and (Bh)
respectively are zero.

For $s_1=2$ one also has
\begin{equation}\label{Nconds2}
\alpha=\beta+\tau=0,
\end{equation}
while for $s_1=1$ (with, as usual, $B$ imaginary) one has
\begin{equation}\label{Nconds3}
\alpha=\beta+\tau. 
\end{equation}

For $\Psi$, (A${}_2$) requires that $\nabla^2\Psi$ satisfies
\begin{eqnarray}\label{A2Wconds}
 (i)~  \nabla^2\Psi_{AB'} &=& 0 ~{\rm for}~ A+B=0,\ldots 5 \nonumber\\
 (ii)~  6\nabla^2\Psi_{51'} &=& \nabla^2\Psi_{60'},\nonumber\\
  (iii)~  5\nabla^2\Psi_{42'} &=& 2\nabla^2\Psi_{51'},\\
 (iv)~  3\nabla^2\Psi_{52'} &=& \nabla^2\Psi_{61'},\nonumber
\end{eqnarray}

In general the conditions on higher derivatives to ensure that
(A${}_m$), $m > 2$, is satisfied do not give new conditions on the
curvature and spin coefficients, so the relevant conditions, which
were used in checking invariances, will not be written out here.

\subsection{The Bianchi identities}
\label{BianImplns}

The Bianchi identities with the canonical choices of frame given in
Section \ref{choices} are written out in Appendix \ref{bianchi}. Some
immediate consequences of those Bianchi identities are collected here.

When all $\Phi_{AB'}=0$, (Bh) and (Bk) give
$\delta\Lambda=0=\Delta\Lambda$ and for LNRI $D\Lambda=0$ so
$\Lambda$ is a constant.

In Segre type [(1,3)] (Bb) shows $\kappa=0$, (Bh) reads
$\delta\Lambda=\rho\Phi_{12'}$ which shows that $\rho$ is real and
(Bg) then implies $\varepsilon=0$. Compatibility with (Bj) then
implies $\sigma=0$. (Bf) shows that $\beta$ is real. Eliminating
$\delta\Phi_{12'}$ and $\Delta\Lambda$ between (Bc), (Bf) and (Bk)
then implies
\begin{equation}\label{abtpident}
  \alpha+\tau+(\pi+\bar{\pi})=\beta.
\end{equation}
Substituting back into (Bf) then shows that $\alpha$, and therefore
$\tau$, are real.

In Segre type [1(1,2)],  (Be) gives
$\rho+\bar{\sigma}=0$, (Bb) shows $\varepsilon$ is real, (Bh) shows
$\alpha-\bar{\beta}$ is real and (Bj) then implies that
$\kappa$ is real. However (Ba) gives $\kappa+\bar{\kappa}=0$, so
$\kappa=0$ and then (Bg) and (Bh) give \eqref{ptident}.
Elimination between (Bh) and (Bj) gives \eqref{lmcond} so
$\lambda+\bar{\mu}$ is a null rotation invariant, and 
\begin{equation}\label{abpident}
  (\alpha+\bar{\alpha})=(\beta+\bar{\beta})+(\tau+\bar{\tau}).
\end{equation}
Hence one can write $\alpha$ in terms of $\beta$ and $\tau$.

% Cf paper I
From \cite{Mac21a}, in Segre type [(11,2)] $\delta\Lambda=0=\kappa$, and
$\Delta\Lambda=\rho=-2(\varepsilon+\bar{\varepsilon})$. In Petrov type
N, $\rho=\epsilon=0=\sigma$ follows from \eqref{Nconds1}, so $\Lambda$
is constant, while in the conformally flat case (Bc) implies
$\sigma=0$ and, setting $\Phi_{22'}=1$,
$\bar{\tau}=2(\alpha-\bar{\beta})$ from (Bd).

%Cf paper II
In Segre type [1(11,1)], (Be) gives $\rho+\bar{\sigma}=0$, (Bb)
implies that $\varepsilon$ is real and (Ba) that $\kappa$ is
imaginary. Then the real parts of $\tau$ and $\pi$ are null rotation
invariants, (Bf) shows that $\mu+\bar{\lambda}$ is real, and (Bg) and
(Bh) show that $\alpha-\bar{\beta}$ is real and \eqref{ptident} holds.
(Bj) and (Bh) imply
$\delta(\Phi_{11'}-\Lambda)=-2(\pi+\bar{\pi})\Phi_{11'}$ and
\eqref{lmcond}. From \eqref{lmcond}, $\alpha$ can be written in terms
of $\beta$ and $\tau$ (or $\pi$). (Bf) and (Bk) yield
$\Delta(\Phi_{11'}-\Lambda)=0$.

\subsection{LNRI and derivatives of $\Phi$}
\label{AmPconds}

For (A${}_1$) the non-zero Ricci curvature components satisfy $DQ=0$
and $\delta Q=\bar{\delta}Q$ as discussed in Section \ref{choices}.
When there is a two-dimensional LNRI group (A${}_1$) further requires
that $\nabla\Phi_{33'}$ and $\nabla^2\Phi_{44'}$ are the only
non-zero components of $\nabla\Phi_{AB'}$ and $\nabla^2\Phi_{AB'}$.

When there is a one-parameter LNRI group (with imaginary $B$),
applying \eqref{zerocond} and \eqref{linearcond} shows (A${}_1$) implies
the following requirements on $\nabla\Phi_{AB'}$:
\begin{eqnarray}\label{A1Pconds}
 (i)~  \nabla\Phi_{AB'} &=& 0  ~{\rm for} ~A+B=0,\ldots 2 \nonumber\\
 (ii)~  3\nabla\Phi_{21'} &=& \nabla\Phi_{30'},\\
    (iii)~  3\nabla\Phi_{22'} &=& 2\nabla\Phi_{31'},\nonumber\\
    (iv)~~ \nabla\Phi_{32'} && {\rm is~ real}.\nonumber
\end{eqnarray}

To complete the study of (A${}_1$) one needs to consider not only
$\nabla\Psi$ and $\nabla\Phi$ but also $\Xi_{AB'}$ (as defined by
$\Xi_{CDEW'}={\nabla^C}_{W'}\Psi_{CDEF}$, \cite{MacAma86}), which is
zero in the conformally flat case. The
conditions on $\Xi_{AB'}$ arising from
\eqref{zerocond}--\eqref{linearcond} are usually trivial to check.

With LNRI of $\Phi$, $\nabla\Phi_{00'}\equiv 0$, (Ba) implies
$\nabla\Phi_{01'}=0$ and (Bb) implies $\nabla\Phi_{02'}=0$, so (i)
above is satisfied. The remaining conditions are best inspected case
by case (see below).

The tests to be applied to $\nabla^2\Phi$ for
(A${}_2$) are as follows.
\begin{eqnarray}\label{A2Pconds}
 (i)~  \nabla^2\Phi_{AB'} &=& 0 ~{\rm for}~ A+B=0,\ldots 3 \nonumber\\
 (ii)~  4\nabla^2\Phi_{31'} &=& \nabla^2\Phi_{40'},\nonumber\\
  (iii)~  3\nabla^2\Phi_{22'} &=& 2\nabla^2\Phi_{31'},\\
 (iv)~  4\nabla^2\Phi_{32'} &=& 2\nabla^2\Phi_{41'},\nonumber\\
    (v)~  4\nabla^2\Phi_{33'} &=& 3\nabla^2\Phi_{42'},\nonumber\\
    (vi)~~ \nabla^2\Phi_{43'} && {\rm is~ real}.\nonumber
\end{eqnarray}

For (A${}_m$), $m >1$, one has to consider, as well as $\nabla^m$
applied to $\Psi_A$, $\Phi_{AB'}$ and $\Lambda$, the derivatives of
order $(m-1)$ of $\Xi_{AB'}$ in Petrov type N, and higher derivatives
of terms computed at previous levels of differentiation (see
\cite{MacAma86}). The required conditions on these for LNRI are
readily inferred from \eqref{zerocond}--\eqref{linearcond}.

\section{LNRI spacetimes} 
\label{LNRIspaces}

The minimal $m$ in (A${}_m$) with LNRI that would ensure an isometry
group as in the criterion (C) quoted in Section \ref{Intro} will now be studied.
In the Petrov type D cases with spatial rotation or boost invariance
treated in the previous papers \citep{Mac21a,Mac21b}, all allowable
Ricci tensors could be treated together. For LNRI spacetimes of Petrov
type N this cannot readily be done and treating the possible Ricci
tensor types separately seems necessary. That also allows one to treat
the conformally flat and Petrov N cases with the same Ricci tensor
type together.

The strategy in doing so is similar to that used in the previous
papers of this series. One first fixes the frame so that the curvature
has an appropriate canonical form, as set out in section
\ref{choices}. Then one can derive the implications of (A${}_1$)
together with the Bianchi identities, as set out in Sections
\ref{AmWconds}--\ref{AmPconds}.  The Bianchi identities, as given in
Appendix \ref{bianchi}, are used together
with the canonically nonzero components of curvature to extract
conditions on the spin coefficients, and the implications of (A${}_1$)
such as \eqref{A1Pconds} are used to find further conditions. The
commutator relations, NP equations and (A${}_m$), $m\geq 2$, then
remain to be considered.

The literature on known metrics for LNRI spacetimes, mostly of
Petrov type N, surveyed in Appendix \ref{metrics}, covers the main
classes of spacetime which can be obtained by assuming (A${}_m$) with
LNRI for sufficiently large $m$. However, more detailed consideration
is needed to find the minimal $m$ for each case that would ensure an
isometry group as in condition (C) in Section \ref{Intro}.

\subsection{LNRI spacetimes with $\Phi_{AB'}=0$}
\label{Nvac}

These are either the well-known conformally flat spacetimes of
constant curvature, which need not be considered further, or are Petrov
type N Einstein spaces.
% File nrnvac.nul
In a canonically chosen frame $\Psi_4$ is constant and the Bianchi
identities imply $\Lambda$ is constant. So $t_0=0$ and, from
\eqref{DPsi}, the only first derivatives of the curvature that could
be nonzero are $\nabla\Psi_{41'}$, $\nabla\Psi_{50'}$ and
$\nabla\Psi_{51'}$. (Bc) gives $4\varepsilon-\rho=0$, so
$\varepsilon=\rho=0$, using \eqref{Nconds1}, and (Bd) gives
$4\beta-\tau=0$.

If $s_1=2$, these and \eqref{Nconds2} imply
\begin{equation}\label{NvacValues}
  \alpha=\beta=\varepsilon=\kappa=\rho=\sigma=\tau=0.
\end{equation}
In this case $k_{a;b}=-(\gamma+\bar{\gamma})k_ak_b$, so $\npk$ is
proportional to (but not necessarily equal to) a covariantly constant
null vector\footnote{In general $k_{a;b}$ is
  $(\alpha+\bar{\beta})k_am_b+\tau \bar{m}_ak_b -
  \rho\bar{m}_am_b-\sigma\bar{m}_a\bar{m}_b+\kappa \bar{m}_al_b +(CC)
  -(\gamma+\bar{\gamma})k_ak_b-(\varepsilon+\bar{\varepsilon})k_al_b$,
  where $(CC)$ stands for the complex conjugate of all the preceding
  terms.}. Thus the spacetimes are $pp$-waves, \eqref{ppwave} (see
\cite{SteKraMac03}, Section 24.5).

With \eqref{NvacValues}, (NPl) shows that $\Lambda=0$. Inspection of
\eqref{DPsi2} shows that the only possible functions of position in
the first derivatives are the real and imaginary parts of $\gamma$.
The only component among the minimal set of second derivative terms
defined in \cite{MacAma86} which is not identically zero is
$\nabla^2\Psi_{62'}= (4\Delta\gamma +4\gamma\bar{\gamma}+20
\gamma^2)\Psi_4$, and similarly at the third derivative only
$\nabla^3\Psi_{73'}$, which is a combination of $\gamma$, its complex
conjugate and $\Delta$ derivatives thereof, can be nonzero.  So
(A${}_1$) implies (A${}_3$), whence $t_1\leq 1$ so there can be at most
one independent function in $\gamma$, say $w$.  If $w$ is constant the
\CKP\ terminates at the first step with $s_1=2$, $t_1=0$ and there is
a group $G_6$. These are (locally) the well-known homogeneous plane
wave spacetimes ((12.12) in \cite{SteKraMac03}). Otherwise $t_1=1=t_2$,
$s_2=2$ and the \CKP\ terminates at step 2.  These are (locally) the
$pp$-waves in which there is a $G_5$ transitive on three-dimensional
null surfaces ((24.46) in \cite{SteKraMac03}).

Thus Petrov type N spacetimes with $\Phi_{AB'}=0$ admitting a
two-dimensional LNRI group are $pp$-waves (including plane waves), and
(A${}_1$) is a sufficient criterion.

% File nrnva1.nul (and log)
Now consider $s=1$ (with imaginary parameter $B$). From (Bc), (Bd) and
the results of Section \ref{AmWconds}  
\begin{equation}\label{NRNs1conds}
\kappa=\sigma=\rho=\epsilon=0,\quad \tau=4\beta,  \quad \alpha = 5\beta.
\end{equation}
If $\beta=0$ the spacetime will have $s=2$ so $\beta\neq 0$ if
$s=1$.
% Because s=2 at steps 1--3 provided glmident holds (see log)
% File nrnva1.red
(NPp), (NPq) and (NPl) read respectively
\begin{eqnarray}
4\delta\beta &=&20 \beta^2-20\beta\bar{\beta},\nonumber\\
4\bar{\delta}\beta &=&20 \beta^2+12\beta\bar{\beta}+2\Lambda,\\
5\delta \beta -\bar{\delta}\beta &=&26\beta\bar{\beta}-10\beta^2+\Lambda,\nonumber
\end{eqnarray}
which imply $\Lambda=20\beta^2-36\beta\bar{\beta}$. Thus $\beta^2$
must be real, so $\beta$ is either real or pure imaginary, and in
either case since $\Lambda$ is constant $\beta$ is constant.  Thus the
only possible independent function(s) of position in the first
derivatives of the curvature tensor are in $\gamma$.  With
\eqref{NRNs1conds}, $\gamma$ is an invariant so $D\gamma=0$. Then
(NPf) implies $\beta(\pi+\bar{\pi})=-8(\beta\bar{\beta})$, which
implies $\beta$ is real, and $(\pi+\bar{\pi})=-8\beta$. Then
\eqref{lmcond} and \eqref{ptident} are true and $\lambda+\bar{\mu}$
is an invariant.

(NPo) and (NPr) give
\begin{eqnarray}
\delta\gamma &=& \beta(-3\gamma
    +\bar{\gamma}+5(\mu+\bar{\lambda})),\nonumber \\
\bar{\delta}\gamma &=& \beta(3\gamma-5\bar{\gamma} +5(\lambda +
   \bar{\mu})).\nonumber
\end{eqnarray}
Since $\gamma$ is an invariant, $\delta\gamma=\bar{\delta}\gamma$ so
$6\gamma+5(\lambda+\bar{\mu})$ is real.  With this and earlier
information, (A${}_2$) can be checked.  Computing $\nabla^2\Psi$ one
finds \eqref{A2Wconds} (i)--(iii) are satisfied, but \eqref{A2Wconds}
(iv) requires $\gamma$ to be real, so $(\mu+\bar{\lambda})$ is also
real, and \eqref{glmident} is satisfied whence $\nu+\bar{\nu}$ is an
invariant. $\nabla^2\Psi_{62'}=
[-20(\nu+\bar{\nu})+24\gamma^2+\Delta\gamma]\Psi_4$, whence
$\nu+\bar{\nu}$ could be a further independent function of position.

Continuing to the further derivatives, using the above information,
shows (A${}_3$) and (A${}_4$) are satisfied.  Thus (A${}_2$) ensures a
one-dimensional null rotation isotropy group in Petrov type N if
$\Phi_{AB'}=0$, but unless a different way to prove $\gamma$ is real
has been overlooked, (A${}_1$) is insufficient. The spacetime has a
$G_3$ if $t_p=2$, or possibly a $G_r$, $r>3$ and is either in the
\eqref{Barnes} or \eqref{siklos} classes. As set out in Appendix
\ref{metrics}, there are known LNRI Petrov N Einstein spaces in these
classes, including the Kaigorodov metric with a $G_5$.

\subsection{LNRI spacetimes with a Ricci tensor of Segre type [(1,3)]}
\label{NRN13}

% File lnri13.nul (and log)
% File cf13.nul for checks in CF case
Using the implications found in Section \ref{BianImplns} from the
Bianchi identities, and  
applying (A${}_1$) to $\nabla\Phi$ one finds that \eqref{A1Pconds} (i)
is true, (ii) implies $\rho=0$, and (iii) together with
\eqref{abtpident} implies \eqref{ptident}, \eqref{lmcond} and
$\alpha=\beta + \tau$. \eqref{A1Pconds} (iv) implies that
$2\gamma+\bar{\lambda}+\mu$ is real, i.e.\ \eqref{glmident} is
true. In the conformally flat case (Bd) implies
$\gamma+\lambda+\bar{\mu}$ is real so $\gamma$ and $\lambda+\bar{\mu}$
are both real.
    
Hence, whether $\Psi_4\neq 0$ or not, \eqref{SCinvce} holds if (A${}_1$) is
true; $\alpha$, $\beta$ and $\tau$ are real; and $\alpha$, $\beta$,
$\gamma$, $\tau$, $\lambda+\bar{\mu}$ and $\nu+\bar{\nu}$ are
invariants. (Bc) now gives $\delta\Phi_{12'}=-2\beta\Phi_{12'}$ and
(Bf) that $\Delta\Lambda=(\pi+\bar{\pi})\Phi_{12'}$.  Note that in the
conformally flat case there is still the freedom to boost to a frame
such that $|\Phi_{12'}|= 1$.

(NPp) gives $\delta\tau=0$ and then (NPf), (NPl) and (NPq) each yield
$\Lambda=-\tau^2$.  If $\tau=0$ then \eqref{ptident} implies
$\pi+\bar{\pi}=0$. Otherwise $-2\tau\Delta\tau = \Delta \Lambda
=(\pi+\bar{\pi})\Phi_{12'} = -2\tau\Phi_{12'}$ implies
$\Delta\tau=\Phi_{12'}\neq 0$, $\tau$ is a non-constant invariant, and
$t_p\geq 1$. 

% File lnr130.nul for tau=0. Nothing in the checks there is upset if Psi_4=0.
(A${}_2$) and (A${}_3$) are now identically fulfilled in all cases above
as a result of the constraints found, regardless of the values
of $\Psi_4$ and $\tau$. Since $s_i=1$ for all $i$ the spacetime must
have at least a $G_3$. Of the metrics in Appendix \ref{metrics} only
\eqref{Barnes}, with a $G_3$ on $N_3$, can support a Ricci tensor of
Segre type [(1,3)], and an example with this Segre type is given there.

\subsection{LNRI spacetimes with a Ricci tensor of Segre type [1(1,2)]}
\label{NRN112}

% File cf112.nul and lrn112.nul
The calculations are a little more complicated in the conformally flat
case, so this is treated first, using the results already stated in Section
\ref{BianImplns}. (Bd) implies that
$\bar{\tau}-2\bar{\beta}-2\alpha$ is real so using the reality of
$\alpha-\bar{\beta}$,
$\tau-\bar{\tau}=4(\beta-\bar{\beta})$.  Adding the imaginary parts of
(Bc) and (Bf) gives \eqref{glmident}.
%test7
On substituting for $\Delta \Phi_{11'}$ and
$\Delta \Lambda$ from the real parts of (Bc) and (Bf) into (Bk) one
obtains $\bar{\rho}+\rho+4\varepsilon=0$. %test8

Now let (A${}_1$) be imposed. \eqref{A1Pconds} (i) and (ii) are
identically satisfied. Evaluating the imaginary part of
\eqref{A1Pconds} (iii), using \eqref{glmident}, implies that $\gamma$
and $\bar{\lambda}+\mu$ are real. Now (Bf) implies $\rho$ is real. The
real part of \eqref{A1Pconds} (iii) gives
$2\rho+\varepsilon=0$ so $\rho=\varepsilon=0$,
and therefore $\sigma=0$.
% ReA1P3 and IMA1P3
Evaluating \eqref{A1Pconds} (iv) then shows that $\beta$ is real and
hence $\alpha$ and $\tau$ are also real and \eqref{abpident}
becomes $\alpha=\beta+\tau$. So \eqref{SCinvce} holds. With these
conditions  (A${}_2$) and (A${}_3$) hold and the spacetime has a group
$G_3$, $r\geq 3$.

From (NPp) and (NPq), $\delta\tau=2\Phi_{11'}=2(\tau^2+\Lambda)$ so
$\tau$ cannot be constant (and hence $t_p\geq 1$).  Since
$\delta(\Phi_{11'}-\Lambda)=4\tau\Phi_{11'}\neq 0$, $t_0 \geq 1$.
Thus $t_1$ could be 1; then the \CKP\ terminates at the first step and
there is a $G_4$, which might allow a metric of the form \eqref{T3G4},
while if $t_p=2$, the \CKP\ stops at step 2 or 3, giving a Barnes
metric, \eqref{Barnes}.  One can also infer from (NPo) and (NPr) that
$\Delta\tau=\Delta(\alpha-\beta)=0$.
% lnr112 checks A2. A3W is OK if delta = bar delta for beta , tau and
%  gamma and D(nu + bar nu) =0. Sim A3Pn see notes.

In the Petrov N case, since $\sigma=0=\rho+\varepsilon$ from
\eqref{Nconds1}, (Be) implies $\rho=\varepsilon=0$.  \eqref{A1Pconds}
will give the same equations as when $\Psi_4=0$, so again $\alpha$,
$\beta$ and $\tau$ are all real and $\alpha=\beta+\tau$, $\gamma$ is
real and so is $\bar{\lambda}+\mu$, and all these are invariants.
These conditions are useful in checking (A${}_2$), which is now easily
found (with CLASSI's help) to be satisfied, as is (A${}_3$). The
deductions in the previous paragraph apply to the Petrov type N case
also.  An example is given in Appendix \ref{metrics}.
 
\subsection{Petrov N LNRI spacetimes with a Ricci tensor of Segre type [(11,2)]}
\label{NRN2}

Although this Segre type of Ricci tensor is invariant under a
three-dimensional group, these spacetimes have only a two-dimensional
LNRI group at the initial step of the \CKP, i.e.\ $s_0=2$, since the
spatial rotation invariance of $\Phi_{AB'}$ cannot be shared by the
Weyl tensor.  The calculations are somewhat similar to those for the
Petrov type N Einstein spaces above. With the usual choice of frame,
\eqref{Nconds1} and \eqref{Nconds3} hold and if $s_1=2$ so does
\eqref{Nconds2}. (Bc) then implies $\varepsilon=\rho=0$, so if $s=1$,
\eqref{SCinvce} is true.  Note that $\alpha$, $\beta$, $\gamma$,
$\tau$, and $\pi+\bar{\pi}$ are invariants whether $s=1$ or 2.

%File nrn22.nul
When $s=2$, so $\delta\Phi_{22'}=0$, then \eqref{Nconds2} and
the Bianchi identity not yet used, (Bd), give
\begin{equation}\label{N22Bd}
5\beta\Psi_4=3\bar{\beta}\Phi_{22'}.
\end{equation}

% \subsubsection{$s=2$, $\beta\neq 0$}
% File nrn220.nul
{\bf If $\bm{s=2}$ and $\bm{\beta\neq 0}$},
taking moduli in \eqref{N22Bd} yields that $\Phi_{22'}= \pm 5/3$.
(NPl) and (NPq) give $\Lambda=-\beta\bar{\beta}\leq 0$. Since, as in
Section \ref{BianImplns} $\Lambda$ is constant, $|\beta|$ is constant
and $t_0=0$. A constant spatial rotation could be used to make
$\beta$ real.
Since $\gamma$ is a null rotation invariant, $\delta\gamma=0$ and (NPo) implies
$\gamma=0$. Inspection shows that $t_1=0$ and the \CKP\ terminates at
the first step. The resulting spacetime is the solution of
\cite{Def69} admitting a $G_6$, which appears as (12.6) in
\cite{SteKraMac03}.
% File defris.spi
(In \cite{SteKraMac03} this metric was arrived
at by assuming the $G_6$ from the start, but the arguments are closely
analogous.)

{\bf If $\bm{s=2}$ and $\bm{\beta=0}$}, \eqref{NvacValues} holds and
as in Section \ref{Nvac} the spacetimes are $pp$-waves. Following the
arguments in that Section, $\Lambda=0$, and as $\gamma$ is a null
rotation invariant, its only possible nonzero derivative is
$\Delta\gamma$. Unless $t_0=0$ and $t_1=1$ the \CKP\ terminates at
step 1, and otherwise at step 2, with respectively a $G_5$ if $t_1=1$
or a $G_6$ if $t_1=0$. The possible metrics have been discussed by
\cite{SipGoe86} (see Appendix \ref{metrics}). For these cases only
(A${}_1$) was needed.

If $s=1$, (NPp) and (NPq) imply $\Lambda=-\tau\bar{\tau}$ so
$|\tau|$ is constant, as in the case $s=2$ with
$\beta\neq 0=\beta+\tau$.  Substituting into (NPl),
remembering that $\beta$ is a null rotation invariant, shows that
$\alpha(2(\alpha-\bar{\alpha})-(\tau-\bar{\tau}))=0$. Hence two cases
result, depending on whether $\alpha=0=\beta+\tau$ or not.

% File nrn221.nul
{\bf If $\bm{s=1}$ and $\bm{\alpha=0=\beta+\tau}$}, then
$\delta\Phi_{22'}=0$ would imply $s_1=2$ by inspection of the only
non-zero quantities with $q=1$ in the minimal set of
\cite{MacAma86}. One also has \eqref{N22Bd}, and by the same arguments
as above one arrives at the Defrise metric, \eqref{Defrise}, or
$pp$-waves.  So one must have $\delta\Phi_{22'}\neq 0$, whence $t_0 =
1$ ($\Psi_4$ and $\Lambda$ being constant), and (A${}_4$) will not be
required. To consider (A${}_2$), note that from (NPp), $\delta
\tau=0=\delta\beta$, whence \eqref{A2Wconds} is satisfied. From
\eqref{A2Pconds} (v) with $\delta\Phi_{22'}\neq 0$,
$2\bar{\beta}=\pi+\bar{\pi}$ so $\beta$ and $\tau$ are real and
constant.  \eqref{ptident} and \eqref{lmcond} are true. If $\beta\neq
0 \neq \Lambda$, (NPo) and (NPr) imply $\gamma$ is
real. \eqref{A2Pconds} (vi) implies (after use of the
$[\delta,\,\Delta]$ commutator) that $\lambda+\bar{\mu}$ is also
real. Then (A${}_3$) is true and there is at least a $G_3$, assuming
(A${}_2$).

% File nrn222.nul
{\bf If $\bm{s=1}$ and $\bm{\alpha\neq 0}$}, then
$(\tau-\bar{\tau})=-2(\beta-\bar{\beta})$ and
$(\alpha-\bar{\alpha})=(\tau-\bar{\tau})+(\beta-\bar{\beta})
   =-(\beta-\bar{\beta})$.
(A${}_1$) is then satisfied.  Since $\gamma$ is an invariant, (NPf)
implies \eqref{ptident} (and thus \eqref{lmcond}).  Checking
(A${}_2$), \eqref{A2Wconds} (ii) and (iii) are then satisfied.
\eqref{A2Wconds} (iv) gives \eqref{glmident}. Conditions
\eqref{A2Pconds} are then satisfied. The quantities $\gamma$,
$\lambda+\bar{\mu}$ and $\nu+\bar{\nu}$ are invariants. Using this
information, (A${}_3$) is then satisfied. The curvature components other than
$\Phi_{22'}$ are constants, so $t_0\leq 1$ while $s_0=2$ and
$s_1=1$. If $\Phi_{22'}$ is not constant, then since $t_2\leq 2$, the
\CKP\ will terminate at the latest at $p=2$ (with a sequence $(2,\,1)
\rightarrow (1,\,1) \rightarrow (1,\,2)$) and there will be a $G_3$ or, if
$t_2=t_1=1$, a $G_4$.  Only if $\Phi_{22'}$ is constant and $t_0=t_1=0$ is 
a check of (A${}_4$) required. The inferences above from (A${}_1$) and (A${}_2$)
remain true and $t_1=0$ implies that (from \eqref{DPsi}) that $\alpha$
and $\gamma$ are constants as well as $\Phi_{22'}$.
% File nrn223.nul
With these extra conditions (A${}_4$) is readily checked (using
CLASSI).

\subsection{Conformally flat LNRI spacetimes with Ricci tensor of
  Segre type [(11,2)]}
\label{CFNRN2}

By Theorem 37.19 of \cite{SteKraMac03}, all conformally flat pure
radiation spacetimes (with $\Lambda=0$) are included in the metrics
(37.104)--(37.106) there. Here the $m$ required in (A${}_m$) to ensure
that an LNRI spacetime (allowing $\Lambda\neq 0$) has null rotation
isotropy is considered. Direct calculation shows that imposing null
rotation invariance on (37.106) of \cite{SteKraMac03} leads (from
\eqref{A2Pconds} (v)) to $\Phi_{22'}=0$, i.e.\ this case will not
arise. With a zero Weyl tensor there is still full boost and spatial
rotation freedom: the boost can and will be used to set $\Phi_{22'}=\pm 1$, and
if $s=1$ the rotation will be used to make $B$ imaginary.

% File cfpure.nul
Summarizing earlier arguments, (Bc) gives $\sigma=0$, (Bg, h and i)
give $\kappa=0=\delta\Lambda$, (Bf) and (Bk) imply that $\rho$ is
real, $\Delta\Lambda =\rho\Phi_{22'}$ and
$\rho=-2(\varepsilon+\bar{\varepsilon})$. \eqref{A1Pconds} (iii) then
implies $\varepsilon+\bar{\varepsilon}=0=\rho$ (so $\varepsilon$ is
imaginary and $\tau$ is an invariant), and $\Lambda$ is constant. Thus $t_0=0$.
(Bd) implies that $\bar{\tau}=2\bar{\beta}+2\alpha$, while
\eqref{A1Pconds} (iv) requires $\tau+\bar{\alpha}+\beta$ to be
real. Thus $\tau=2(\alpha+\bar{\beta})$ is real. The imaginary part of
(NPp) now implies
$\tau=0$ or $\beta-\bar{\alpha}$ is real, so $\alpha$ and $\beta$ are
both real.

{\bf If $\bm{\tau=\alpha+\bar{\beta} = 0}$}, $\npk$ is proportional to a
covariantly constant vector, as in Section \ref{Nvac}, and the
spacetime is a $pp$-wave with a $G_5$. These cases have a
2-dimensional LNRI group and are LRS so $s_1=3$. They are the
solutions (37.104)–(37.105) in \cite{SteKraMac03}.

{\bf If $\bm{\tau\neq 0}$}, $\alpha$, $\beta$ and $\tau$ are all real and
$\alpha+\beta\neq 0$. Then (NPq)
gives $\Lambda=-2(\alpha^2-\beta^2)$.  The imaginary part of (NPc)
implies $\varepsilon - \bar{\varepsilon}=0$ so $\varepsilon$ is zero
and $\alpha$ and $\beta$ are invariants. Then \eqref{A1Pconds} is
satisfied. Since $\nabla\Phi_{23'}=\tau\Phi_{22'} =\pm \tau \neq 0$, 
$s_1=1=s$. 

Checking (A${}_2$),
\eqref{A2Pconds} (i)--(iv) are automatically satisfied.
\eqref{A2Pconds} (v) gives $\alpha+3\beta=0$ so $\alpha= -3\beta$,
$\tau=-4\beta$ and $\Lambda=-\tau^2 \neq 0$. Thus $\tau$, $\alpha$ and
$\beta$ are constants. \eqref{Nconds3} holds so $\gamma$ is an
invariant. (NPf) now gives \eqref{ptident} or equivalently (in this
case) \eqref{lmcond}, and \eqref{A2Pconds} (vi) gives \eqref{glmident}. So
$\lambda+\bar{\mu}$ and $\nu+\bar{\nu}$ are invariants.  Using
\eqref{glmident}, (NPo) and
(NPr) now imply $\gamma$ and $\lambda+\bar{\mu}$ are real,
and $\delta\gamma = -2\beta\gamma-3\beta(\lambda+\bar{\mu})$.
(A${}_3$) is then satisfied.

% File cfpur4.nul
Since $s_0=3$ and $s_1=1$, conditions arising from (A${}_4$) might be
required to prove null rotation isotropy if $t_1$ were zero. This
would require $\gamma$ to be constant and $\lambda+\bar{\mu}$ to be
zero. With these conditions, (A${}_4$) is readily checked (using
CLASSI). If $\gamma$ is not constant, $t_1=1$. Among the second
derivatives, only
$\nabla^2\Phi_{44'}=12(\nu+\bar{\nu})+24\gamma^2+\Delta\gamma$ could
contain another function of position. If it does, the \CKP\ terminates
with $p=2$ and $t_p=2$ and otherwise with $p=1$ and $t_p=1$.

\subsection{LNRI spacetimes with Ricci tensor of Segre type [1(11,1)]}
\label{NRN111}

% File nr1111.nul
In the Petrov type N case, the relations given in Section
\ref{BianImplns}, and \eqref{Nconds1} and \eqref{Nconds3}, which
follow from (A${}_1$), together give \eqref{SCinvce}.  Hence whatever
the value of $\Psi_4$, $\varepsilon\Psi_4=0$ and (Bc) then shows
$\gamma$ is real. (Bg), (Bh) and (Bj) show $\alpha-\bar{\beta}$ is
real and \eqref{ptident} and \eqref{lmcond} are true. (Bf) implies
$\lambda+\bar{\mu}$ is real. Thus \eqref{glmident} is true and
$\lambda+\bar{\mu}$ and $\nu+\bar{\nu}$ are invariants.  Reality of
$\alpha-\bar{\beta}$ and \eqref{Nconds3} give
$\tau-\bar{\tau}=-2(\beta-\bar{\beta})=2(\alpha-\bar{\alpha})$. (A${}_1$),
(A${}_2$) and (A${}_3$) are then satisfied. As $s_0=3$ while $s_1=1$,
one must check if (A${}_4$) is needed to ensure null rotation
isotropy.
% File nr1110.nul
However, the conditions for $t_0=t_1=0$ are very strong
($\Lambda$, $\Phi_{11'}$, $\alpha$, $\gamma$, $\nu+\bar{\nu}$,
$\tau+\bar{\tau}$ and $\beta+\bar{\beta}$ must be constants and
$\lambda+\bar{\mu}=0$), and with these conditions (A${}_4$) is
satisfied. The spacetime has at least a $G_3$.

% File cf1111.nul
In the conformally flat case, (A${}_1$) is again satisfied, as a
consequence of the Bianchi identities, and from (Bd) $\nu+\bar{\nu}=
0$. \eqref{A2Pconds} (i) and (vi) are then satisfied.

{\bf If $\bm{(\mu+\bar{\lambda}) \neq 0}$}, then \eqref{A2Pconds} (ii)
and (iii) imply that $\rho$ is real and $\varepsilon+2\rho=0$, while
\eqref{A2Pconds} (iv) implies that $2\beta+\tau$ is real. As
$\alpha-\bar{\beta}$ is invariant, the imaginary part of (NPd) added
to the conjugate of (NPe) yields that $\kappa=0$. Then $\rho$ is a
null rotation invariant, so $D\rho=0$ in (NPa) which implies
$\rho=\sigma=\varepsilon=0$. Then (A${}_3$) is satisfied, and since
$t_0\geq 1$ the \CKP\ must terminate at $p=1$ or $p=2$. (A${}_2$)
implies null rotation isotropy in this case.

% File cf1110.nul
{\bf If $\bm{(\mu+\bar{\lambda})=0}$}, no information arises from
(A${}_2$).  In this case, from (Bc, f and k)
$\Delta\Lambda=\Delta\Phi_{11'}=0$.  The only spin coefficient
combination appearing in the symmetrized spinor derivatives of $\Phi$
is $\tau+\bar{\tau}$ and its derivatives and
$\delta(\Phi_{11'}-\Lambda)=(\tau+\bar{\tau})\Phi_{11'}$ (from (Bg, h
and j)). Applying the $[\delta,\,\Delta]$ commutator to
$(\Phi_{11'}-\Lambda)$ gives $\Delta(\tau+\bar{\tau})=0$.  (A${}_2$)
and (A${}_3$) are then satisfied. If $t_0=t_1=0$, so $\Phi_{11'}$,
$\Lambda$ and $\tau+\bar{\tau}$ are constant, (A${}_4$) could be
required to ensure the \CKP\ terminates at $p \leq 3$, but inspection
shows that then $t_2=0$ also, so the \CKP\ will terminate with
$p=1$. In this subcase only (A${}_1$) has been used, and only for directional
derivatives of $\Phi_{11'}$ and $\Lambda$.

\section{Conclusion}

In this paper, it has been shown that the
conjecture of Siklos, although not correct in general, as discussed in
the introduction to Paper 1 of this series \citep{Mac21a}, is correct
in almost all cases with local null rotation invariance. Thus the following holds:

\begin{theorem}
  \label{thm1}
Einstein spacetimes with a Weyl tensor of Petrov type N, spacetimes
with a Ricci tensor of Segre type $[(11,2)]$, and conformally flat
spacetimes with a Ricci tensor of Segre type $[1(11,1)]$, are locally
null rotation isotropic if and only if their curvature tensor and its
first two derivatives are locally null rotation invariant.
\end{theorem}

As explained in Section \ref{sec:LNRIImplns}, these spacetimes might 
have exhibited a sequence of $(s_i,\,t_i)$ values requiring a check of
the fourth derivatives of the curvature, but the detailed calculations
show that this is never the case. 

\begin{theorem} A spacetime other than those covered by Theorem
  \ref{thm1} is locally null rotation isotropic if and only if the
  curvature tensor and its first derivative are locally null rotation
  invariant.
\end{theorem}

All locally null rotation isotropic spacetimes admit a
multiply-transitive $G_r ~(r\geq 3)$.

\section*{Acknowledgements}
I am grateful to Jan {\AA}man for his work in developing the software
CLASSI and for discussions of the Petrov type N cases, and to Filipe Mena,
both for the stimulus to this work given by our joint study of discrete
isotropies and for his very careful reading of the first draft of this
paper and useful suggestions for its improvement. I am also grateful
to Charles Torre for correction of an error in that draft, and to
Graham Hall for comments.

\section*{Conflict of interest}

The author is an Editorial Board member of the journal General
Relativity and Gravitation.

\appendix
\section{Known spacetimes with null rotation
  invariance}
\label{metrics}

Local invariance groups $g$ containing both a null rotation and either
rotational or boost symmetry are possible only in conformally flat
spaces with Ricci tensors of Segre types [(11,2)] or [1(11,1)] or
[(111,1)]. Such spacetimes were discussed in the previous papers
\citep{Mac21a,Mac21b}. So the only metrics to be considered in this paper are
those where $g$ consists solely of null rotations.

Spacetimes with a two-dimensional LNRI group have at least a
$G_5$ on an $N_3$. These are included in the $pp$-waves discussed below.
Spacetimes with a one-dimensional LNRI group must have
a group of motions $G_r$ with $r\geq 3$. If $r=3$ the group acts
on a two-dimensional null surface $N_2$. If $r=4$ it acts on a
three-dimensional hypersurface which may be null ($N_3$) or timelike
($T_3$). If $r \geq 5$ but $s=1$ the group acts on the whole spacetime. Groups
$G_6$ or larger which contain null rotation isotropy necessarily act on the
whole spacetime.

The $pp$ waves have the metric form
% File ppwavg.spi
% and for a vacuum solution
% File ppwavv.spi
\begin{equation}\label{ppwave}
  \d a^2 = 2[-\d u \d v - H\d u^2 + \d \xi \d \bar{\xi}], \quad
  H=H(\xi, \bar{\xi}, u);
\end{equation}
which is conformally flat only if $H_{,\bar{\xi}\bar{\xi}}=0$. Of the
energy-momenta considered in \cite{SteKraMac03} only vacuum and pure
radiation (including Einstein-Maxwell) solutions are possible, with
$\Lambda=0$. If $t_p=0$ the solutions are plane waves, which are
special cases of \eqref{ppwave}, and there is a group $G_6$ acting
transitively on spacetime, or in even more special cases with
additional symmetry, a $G_7$ (see (24.46) and Section 12.2 in
\cite{SteKraMac03}).

The plane waves have $s=2$. Expressing this in terms of transitivity
on the set of possible wave surfaces enabled \cite{Hal15,Hal19} to
show the equivalence of two definitions of
the plane wave subcases of \eqref{ppwave}. He also proved that
transitivity on the wave surfaces leads back to the metric \eqref{Defrise}
or those plane waves found in Sections \ref{Nvac}, \ref{NRN2} and
\ref{CFNRN2} above. (See the cited papers for the detailed assumptions
made.) 

% File sigo9.spi to sigo15.spi
The metric \eqref{ppwave} has at least a group $G_1$ generated by
$\npk=\partial_v$, but may admit a $G_r$ with $r=2$ to 7. The possible
symmetry classes were studied by \cite{SipGoe86}, without the use of
the Einstein field equations. Those which have LNRI have
$(s,\,t_p)=(2,\,1)$ and a $G_5$, or $(s,\,t_p)=(2,\,0)$ and a $G_6$, or
$s=1$. There are none with a maximal $G_4$ so $(s,\,t_p)=(1,\,1)$ does
not arise. If $(s,\,t_p)=(1,\,2)$, there is a $G_3$ on $N_2$ so these
are included in \eqref{Barnes} below. If $(s,\,t_p)=(1,\,0)$ there is a
$G_5$.

The $pp$-waves with a $G_5$ or $G_6$ are listed as cases 9 to 15 in
Table II in \cite{SipGoe86}. Case 9 is a homogeneous pure radiation
spacetime, (12.36) in \cite{SteKraMac03}. Case 10 is the only metric
with a $G_5$ on null hyperplanes and $t_p=1$: it is interpretable as
an Einstein-Maxwell plane wave.  Of those with a $G_6$, cases 11-14
are pure radiation plane wave metrics, while case 15 is a vacuum
solution of Einstein's equations. The formulae for these cases appear
in the first paragraph of page 386 of \cite{SteKraMac03}, where
$\kappa$ may be zero.

Note that despite claims to the contrary in the literature, repeated
in \cite{SteKraMac03} Section 24.5, all these $pp$-wave metrics with
LNRI admit interpretations as Einstein-Maxwell spacetimes, although in
one case (that given as (24.47) or equivalently (12.36) in
\cite{SteKraMac03}) the Maxwell field does not inherit the spacetime's
symmetries \citep{Tor12}.

The general metric with a $G_3$ on $N_2$ was given by \cite{Bar79},
who showed that it was equivalent to metric forms used earlier by
Petrov and by Defrise. It can be written
% File barnes.nul
\begin{equation}\label{Barnes}
\d s^2 = B^2(x,\,u)[-2\d u\,\d v +2F(x,\,u)\d u^2 +\d y^2]+\d x^2.
\end{equation}
This has as (possibly) non-zero entries in $\Phi_{AB'}$ exactly the
components allowed in the canonical forms compatible with LNRI in
which $\Phi_{AB'}$ is not identically zero.
In the null
tetrad $\{B(\d v -F \d u),\,B\d u,\,(dx-iB\d y)/\sqrt{2}, (\d x +iB\d
y)/\sqrt{2}\}$ the curvature components are
\begin{eqnarray}
  \label{barnpsi}
\Psi_4 &=& -\half(F_{,xx}+F_{,x}B_{,x}/B -B_{,uu}/B^3+2(B_{,u})^2/B^4),\\
\label{barn11}  
\Phi_{02'}&=&2\Phi_{11'}=-\half(BB_{,xx}-(B_{,x})^2)/B^2,\\
\label{barn12}
\Phi_{12'}&=& -(BB_{,ux}-B_{,u}B_{,x})/\sqrt{2}B^3\\
\label{barn22}
\Phi_{22'}&=& -\half(F_{,xx}+3F_{,x}B_{,x}/B +B_{,uu}/B^3-2(B_{,u})^2/B^4),\\
\label{barnl}
\Lambda &=&-\quarter(BB_{,xx}+(B_{,x})^2)/B^2.
\end{eqnarray}
Boosting from this frame to the frame with $|\Psi_4|=1$ used in the
main text would in general lead to more complicated formulae.  The
$G_3$ is in general maximal but there are cases with a $G_r$, $r>3$,
such as $pp$-waves, and spacetimes with a $G_4$ on $T_3$ \citep{Mac80}.

Only if $B_{,x}=0$ can one have $\Phi_{11'}=\Phi_{02'} = \Lambda=0$ in
the metric \eqref{Barnes}, and then $\Phi_{12'}=0$ and $\Phi_{22'}+\Psi_4=0$.
$\Phi_{12'}=0$ if and only if $B$ is separable, taking the form
$B_1(u)B_2(x)$.  If $\Phi_{11'}=\Phi_{02'}=0$ then $B=Y(u)e^{K(u)x}$
where $Y(u)$ and $K(u)$ are functions of $u$ alone; then
$\Lambda=-K^2/2$.  If $\Lambda=0$ then $B$ has the form
$Y(u)\sqrt{x+K(u)}$, $Y$ and $K$ again being functions of $u$.
Spacetimes obeying one of $\Phi_{11'}=0$, $\Phi_{12'}=0$ or
$\Lambda=0$, with suitable $F$, may also obey $\Phi_{22'}=0$ or
$\Psi_4=0$.

Barnes considered the Einstein equations for \eqref{Barnes} for a
Ricci tensor of Segre type [(112)] (``pure radiation'') with
$\Lambda=0$ and showed that all such spacetimes are $pp$-waves; they admit a
Maxwell field that inherits the LNRI symmetry only in the more special
plane wave case (cf.\ Section \ref{NRN2} above).

Barnes also found an Einstein space with the
$B=Y(u)e^{K(u)x}$ form where $K=C$, a constant, $Y=1/u$ and
$F=G(u)e^{-3Cx}$.
% File barnrl.nul. Use PSUBS
The existence of this Einstein space and others mentioned
below shows that the assertion of \cite{SalGarPle83} that nontwisting
type N spacetimes with a cosmological constant admit at most a $G_2$ of
isometries is mistaken.
% File barnra.nul
% File kaigor.hnl
There are further Petrov type N Einstein spaces of the form
\eqref{Barnes} which Barnes overlooked.  In these, $K=C$ and
$F=G(u)e^{-3Cx}$ again, but $Y(u)$ is constant, say 1. If $G(u)$ is
constant this is Kaigorodov's solution ((12.34) and (38.2) in
\cite{SteKraMac03}), up to constant scale factors, and otherwise is
the $H=A(u)x^3$ solution for \eqref{siklos} below.

One can find further metrics for other Segre type, Weyl tensor, and
$\Lambda$ combinations, though these are generally of less physical
significance. Some examples are, by Segre type:\\
{}[(1,3)], Petrov N, $\Lambda \neq 0$; $B=u^{x-1}$, $F=1$.\\ %bar13.spi
{}[1(1,2)], Petrov N, $\Lambda\neq 0$; $B=ux$, $F=1$. \\ % bar12.spi
{}[1(1,2)], conformally flat, $\Lambda=0$; $B=\sqrt{x}/u$, $F=\sqrt{x}$.\\
{}[(11,2)], Petrov N, $\Lambda\neq 0$; $B=Y(u)e^{Cx}$, $F=1$. \\ % barra.spi
         % Developed from example in Jan's barr.spi
{}[1(11,1)], Petrov N, $\Lambda=0$; $B= \sqrt{x}/u$, $F=1/\sqrt{x}$.

% File siklos.hnl
\cite{Sik85} introduced the class of Lobatchevski plane waves, which
include the only Einstein spaces conformal to $pp$-waves. The metrics
can be written (omitting a constant factor fixing an overall scale for
the curvature) as
\begin{equation}\label{siklos}
  \d s^2 = [-2 \d u \d v - H\d u^2 +(\d x^2+\d y^2)]/x^2, \quad
  H=H(x, y, u).
\end{equation}
They are of Petrov type N and have in general a pure radiation
energy-momentum tensor (Segre type [(11,2)]), a constant negative
$\Lambda$ (so $pp$-waves are not possible), and in general only a $G_1$ of
isometries generated by $\partial_v$. Particular cases may
admit a $G_r$, $2 \leq r \leq 6$: the special $H$ admitting these
larger groups, and their associated Killing
vectors, are listed in Table 1 of Siklos' paper. From that table,
% Case 5 is the general case with LNRI. It has $\Psi_4= x^2H_{,xx}/4$
% $\Phi_{22'}= x^2H_{,xx}/4-xH_{,x}/2$ with scaling so $\lambda=-1/2$
% Cases 9-13 are the special cases with r>3 listed below. 
% File siklo5.hnl
there is a one-parameter LNRI group only if $H_{,y}=0$.  Under that
condition the spacetime admits a $G_3$ on $N_2$ in general, a $G_4$ on
$T_3$ if $H$ is $A(x)$, a $G_4$ on $N_3$ if $A(u)x^2$ or
$u^{-2b-2}A(xu^b)$ with constant $b$, a $G_5$ on $V_4$ if $H=\pm x^a$
for constant $a$ and a $G_6$ on $V_4$ if $H=\pm x^{-2}$. The last of
these is Defrise's pure radiation solution with a cosmological
constant,
(12.6) in \cite{SteKraMac03}, i.e.
\begin{equation}
  \label{Defrise}
  \d s^2 = \frac{3}{|\Lambda|y^2}\left[\d y^2+\d z^2 -\d v\left(\d
      u-\frac{\Lambda\d v}{|\Lambda|y^2}\right)\right].  
\end{equation}
($\Lambda$ here being the cosmological constant, not the Newman-Penrose
$\Lambda$.)  If $H=A(u)x^3$ in these cases the spacetime is an
Einstein space, which when $A$ is constant is Kaigorodov's solution,
(38.2) in \cite{SteKraMac03}. The $G_5$ of the Kaigorodov solution is
maximal for type N Einstein spaces.

The possible metrics for a $G_4$ acting on a null hypersurface $N_3$
(not necessarily with LNRI) were found by Kruchkovich and Petrov. The
list given in \cite{Pet69}, Section 32, has a few readily correctable
misprints. It also includes two cases with non-Lorentzian signature,
and one which (as noted by Petrov) does not allow a null
hypersurface. Inspection of the remaining metrics shows that in most
cases the isotropy in the given isometry group $G_4$ is a spatial
rotation rather than a null rotation.

These remaining metrics were considered further by \cite{LauRay77},
who found no solutions of Einstein's equations for a range of possible
energy-momenta in all but one case, but noted, without finding
general-relativistic solutions, that three metrics admitted additional
Killing vectors.
% Petrov G_4III appears to be Petrov type III with extra KV and Lambda=0
% but this cannot be right as Petrov III does not admit any isotropy. 
% File p3217.hnl
Two of these they denote by $G_4VI_2$ and $G_4VI_4$
in their table (page 885) but $G_4VI_1$ and $G_4VI_2$ on page
886\footnote{A comparison of Petrov's and other enumerations of the
  possible structures of real $G_4$ can be found in \cite{Mac99}.}.
One of those, (32.26) in Petrov's book, where it is labelled
$G_4VI_3$ and also appears as (33.1) labelled $G_5A$, admits a $G_5$
on $N_3$ and is a $pp$-wave in coordinates different from those of
\eqref{ppwave} above. The other, (32.27) or $G_4VI_2$ in the book,
taking $e=1$ to give the correct signature, is a special case of (32.26)
with plane symmetry.

% File p3220.hnl
Lauten and Ray's one relativistic solution is for Petrov's $G_4I_1$
metric, which they give with misprints corrected. It is a pure
radiation solution with an arbitrary $\Phi_{22'}$ dependent on their
$(x_1-x_4)$ and $\Lambda\neq 0$: they call it a null fluid with a
cosmological constant. The metric is a special case of \eqref{Barnes}, as
can be shown using the calculations in Barnes' paper, with an
additional translational Killing vector in the null hyperplane $u={\rm
  constant}$.

The possibilities for a group $G_4$ with LNRI acting on a $T_3$
were studied in \cite{Mac80}, and are either special cases of
\eqref{Barnes} or have metrics of the form
% File sk1319.hnl
\begin{equation}\label{T3G4}
\d s^2 = A^2(z)[ -2e^y\d u(\d v +B(z)e^y \d u) + \d y^2]+\d z^2,
\end{equation}
which appears as (13.19) in \cite{SteKraMac03}. Its Ricci tensor has
Segre type [1(1,2)] or one of its degeneracies, and it is conformally
flat only if $B=0=\Phi_{22'}$ (Segre type [1(11,1)]. In that case
$\Phi_{11'}=(1+4({A_{,z}}^2-AA_{,zz}))/16A^2$.

Finally one could have a homogeneous spacetime with a group $G_5$ acting
transitively and including a null rotation isotropy, when $s=1$ and
$t_p=0$. Such metrics are known for an Einstein space,
for pure radiation, and for a combination of these (see (12.34), (12.36)
and (12.38) in \cite{SteKraMac03}) but these examples are special
cases of the other metrics above.

One can thus anticipate that assuming (A${}_m$) when $g$ contains a
null rotation invariance will lead to spacetimes whose metrics can be
put in one of the forms \eqref{ppwave}, \eqref{Barnes},
\eqref{siklos}, or \eqref{T3G4}. They may have Petrov type N or be
conformally flat, and the Ricci tensors can be: for \eqref{ppwave},
[(11,2)]; for \eqref{Barnes} any of those types considered above; for
Siklos, only [(11,2)] with a non-zero $\Lambda$; and for \eqref{T3G4},
[1(1,2)] or [1(11,1)]. However, with Segre type [1(11,1)] where,
unlike all those metrics, $\kappa$ need not be zero, there may be
other possibilities.

\section{The Bianchi identities under LNRI}
\label{bianchi}

This appendix gives the Bianchi identities assuming, for the reasons
discussed in Section \ref{choices}, that the only components of the
curvature which might be non-zero are $\Psi_4$, which is constant and
such that $|\Psi_4|=1$, $\Lambda$, and the real components
$\Phi_{02'}=\Phi_{11'}$, $\Phi_{12'}$ and $\Phi_{22'}$.

In more detail, only the following components of $\Phi_{AB'}$ remain
in the allowed Segre types: in [(1,3)], $\Phi_{12'}$; in [1(1,2)],
$2\Phi_{11'}=\Phi_{02'}$ and $\Phi_{22'}$; in [(11,2)], $\Phi_{22'}$;
and in [1(11,1)], $2\Phi_{11'}=\Phi_{02'}$. Since $\Phi_{02'}$, if
non-zero, will be equal to $2\Phi_{11'}$ in the canonically chosen
frames, that is assumed below.

For all curvature components, $DQ=0$, and $\delta Q=\bar{\delta}Q$ is
real. Since under these conditions (Be) reads
$(\rho+\bar{\sigma})\Phi_{11'}=0,$ that combination of terms has been
eliminated from (Bb) and (Bi). The conditions \eqref{Nconds1} and
\eqref{Nconds3} have been applied to terms involving $\Psi_4$.

%\newpage
\begin{eqnarray*}
  Ba:& -2(\kappa+\bar{\kappa})\Phi_{11'} =0,\\
  Bb:&  2(2\varepsilon-2\bar{\varepsilon})\Phi_{11'}
          -2\kappa\Phi_{12'}= 0,\\
  Bc:& \delta\Phi_{12'}-2\Delta\Phi_{11'}
     =5\varepsilon\Psi_4 +2(2\gamma-2\bar{\gamma}+\bar{\mu}+\lambda)\Phi_{11'}\\ 
     &             +2(\bar{\tau}-\alpha)\Phi_{12'}-\bar{\sigma}\Phi_{22'}\\       
  Bd:&
  \delta\Phi_{22'}-\Delta\Phi_{12'}=(4\beta-\tau)\Psi_4-2(\nu+\bar{\nu})\Phi_{11'} \\
  &    \qquad\qquad  +2(\gamma+\bar{\mu}+\lambda)\Phi_{12'}
        +(\bar{\tau}-2\bar{\beta}-2\alpha)\Phi_{22'}\\
  Bf:& 2\Delta\Lambda-\delta\Phi_{12'} =
  -2(\mu+\bar{\lambda})\Phi_{11'}+2(\pi+\bar{\pi}+\beta)\Phi_{12'}\\
     & +(\bar{\rho}-2\varepsilon -2\bar{\varepsilon})\Phi_{22'} \\
  Bg:& 2\delta \Phi_{11'}-2\delta \Lambda =
       2(2(\bar{\alpha}-\beta)-(\pi+\bar{\pi}))\Phi_{11'}
         -2(\bar{\rho}-\varepsilon)\Phi_{12'} +\bar{\kappa}\Phi_{22'}\\
Bh:& 2\delta \Phi_{11'}-2\delta \Lambda  =
    2(2(\alpha-\bar{\beta}) + (\tau+\bar{\tau}))\Phi_{11'}-2\rho\Phi_{12'},\\
Bi:& (\kappa+\bar{\kappa})\Phi_{12'}=0,\\
Bj:&-3\delta\Phi_{11'}+3\delta\Lambda = 2(-2\alpha +2\bar{\beta}
                       +\pi+\bar{\pi}-\tau-\bar{\tau})\Phi_{11'}\\
   &\phantom{=-3\delta\Phi_{11'}+3\delta\Lambda=}
     +(\bar{\rho} +\sigma+ 2\rho-2\bar{\varepsilon})\Phi_{12'}-\kappa\Phi_{22'}\\
Bk:& 3\Delta\Lambda-2\delta\Phi_{12'}+\Delta\Phi_{11'}
      = (\rho+\bar{\rho}-2\varepsilon-2\bar{\varepsilon})\Phi_{22'}
             -2(\mu+\bar{\mu}+\lambda+\bar{\lambda})\Phi_{11'}\\
& \phantom{3\delta\Lambda-3\delta\Phi_{12'}=(\rho+\bar{\rho})}
     +(2(\beta+\bar{\beta}+\pi+\bar{\pi})-\tau-\bar{\tau})\Phi_{12'}
\end{eqnarray*}

% \bibliography{GRstrings,isotropy}
% For now use my ApA-like. Amend for publication
% \bibliographystyle{mapalike}

\end{document}